# Low time resolution analysis of polar ice cores cannot detect impulsive nitrate events

D.F. Smart[1], M.A. Shea[1], A. L. Melott[2], and C. M. Laird[3]

1. 100 Tennyson Avenue, Nashua, New Hampshire 03062 USA email sssrc@msn.com
2. Department of Physics and Astronomy, University of Kansas, Lawrence, Kansas 66045 USA email melott@ku.edu
3. 1010 E. 450 Rd, Lawrence, Kansas, 66047 email claude.m.laird@gmail.com

**Abstract**

Ice cores are archives of climate change and possibly large solar proton events (SPEs). Wolff et al. (2012) used a single event, a nitrate peak in the GISP2-H core, which McCracken et al. (2001a) time associated with the poorly quantified 1859 Carrington event, to discredit SPE-produced, impulsive nitrate deposition in polar ice. This is not the ideal test case. We critique the Wolff et al. analysis and demonstrate that the data they used cannot detect impulsive nitrate events because of resolution limitations. We suggest re-examination of the top of the Greenland ice sheet at key intervals over the last two millennia with attention to fine resolution and replicate sampling of multiple species. This will allow further insight into polar depositional processes on a sub-seasonal scale, including atmospheric sources, transport mechanisms to the ice sheet, post-depositional interactions, and a potential SPE association.

## 1.    Introduction

Atmospheric ionization events break $N_2$ chemical bonds, enabling nitrogen oxides [NO(y)] formation, normally at low atmospheric abundance [e.g. *Thomas et al.,* 2005]. There has been controversy whether NO(y) deposition in ice cores, primarily as nitrate ($NO_3$), can be used as a proxy for very large solar proton events (SPEs). Based on the impulsive nitrate events identified in polar ice by *Dreschhoff and Zeller* [1990, 1994, 1998] *McCracken et al.* [2001a] established a calibration between impulsive nitrate deposition in the 415-year GISP2-H ice core obtained from the top of the ice sheet divide at Summit, Greenland, in 1991 by *Dreschhoff and Zeller* [1994, 1998] and extremely large SPEs between 1956 and 1989.  These authors identified 70 impulsive nitrate events in the GISP2-H core and concluded that they could be used as proxies to identify past extremely large SPEs. One of the largest impulsive nitrate events was dated as late 1859 and associated with the well-documented but poorly quantified September 1859 Carrington event [*Clauer and Siscoe*, 2006].  *Wolff et al.* [2012] analyzed several central Greenland ice cores, including their D4 [*McConnell et al.*, 2007] and Zoe [*Wolff et al.*, 2012] cores for a 40-year time period (1840-1880) and suggested that the impulsive nitrate event identified by *McCracken et al.* [2001a] as the 1859 Carrington event most likely resulted from biomass burning in 1863.

It is the hard spectra solar cosmic ray ground-level events that have sufficient >300 MeV fluence to penetrate to and through the ozone layer (30 to 17 km altitude).  The large fluence of these particles produces the ionization that provides the energy to ionize $N_2$ in the lower stratosphere or even the troposphere [*Atri* et al., 2010; *Usoskin et al.,* 2011].  When NO(y) is generated at these lower altitudes, there is the possibility of moving quickly into the troposphere, either by downward advection or by incorporation into polar stratospheric clouds and  gravitational settling, where it will be deposited as NO(y) in snow events.  The softer spectra events will deposit the ionization energy at higher altitudes, and the resulting NO(y) must be transported downward by mechanisms such as the upper branch of the Brewer-Dobson circulation.  These well-established transport mechanisms acting over longer time periods should significantly broaden the deposition timescale as the material diffuses down into the troposphere making any nitrate signal in the ice much harder to identify.

The problem with using the Carrington event to dispute all SPE effects in polar ice nitrate profiles as done by *Wolff et al.* [2012] is that while this was clearly a major solar event (white light central meridian solar flare) with subsequent associated extreme geomagnetic storm and low-latitude aurora, it is only a single example and little quantitative information exists with respect to a SPE.  Hence, it may not represent the type of high-energy SPE needed to produce an impulsive nitrate event and is not the ideal test case.  A careful inspection of Figure 1 in *McCracken et al.* [2001a] shows a complex structure of several events occurring in the 1859 dated GISP2-H core, which is not seen in the 1863 Zoe and D4 data.  It is simply not possible at this time to state conclusively whether the large nitrate spike dated to 1859 or 1860 in the GISP2-H core and to 1862 or 1863 in other Greenland cores resulted from the Carrington event or if a smaller, or no, impulsive nitrate enhancement should be expected.

In an effort to understand the different interpretations between *McCracken et al.* [2001a] and *Wolff et al.* [2012], we concluded that a correlation test between impulsive nitrate spikes in polar ice and SPEs could be conducted by examining ice core nitrate deposition during the only years when both well-documented extremely large hard spectrum relativistic solar cosmic ray events occurred and ultrafine-resolution nitrate measurements in consolidated firn exist, namely the 14-year interval 1937 to 1951. We use this time period to ascertain if the ice core data available for analysis contain evidence of short-duration impulsive nitrate enhancements that could be time associated with these large relativistic solar cosmic ray events. Using spectral analysis we also determine the resolutions of sampling schemes using these nitrate enhancements. In this study we will show that finer than 0.08 year (sub-monthly) resolution is necessary for impulsive nitrate event detection in ice cores and conclude that the data used by *Wolff et al.* [2012] were inadequate to differentiate short-duration nitrate enhancements that might be indicative of an extremely large solar proton event.

We note that *Schrijver et al.* [2012] also discuss reasons to question the validity of nitrate spikes. We comment briefly on those here. (1) The conventional nitrate transport processes from the middle stratosphere and above is very slow, so it is hard to make spikes in the ice. The problem with this is an assumption that substantial deposition cannot occur at lower altitudes. This can be shown to be incorrect by examining tables in *Atri et al.* [2010] or balloon data in *Nicoll and Harrison* [2014]. (2) Nitrate interference from biomass burning is a serious concern but can be filtered out by examining other species (e.g., black carbon, ammonium and vanillic acid), at comparable resolution. Similarly, post-depositional variations could be detectable by examining the concentration of other species. Ionization events should only involve impulsive nitrate increases although conductivity may increase also as nitrate is highly ionic in water and is the second largest contributor to Greenland ice conductivity after sulfate.

## 2.    What is an "Impulsive Nitrate Event"?

There appears to be disagreement as to what constitutes an "impulsive nitrate event" in the ice core data. In fact, for years, these features either were not observed due to insufficient time resolution or were attributed to contamination [e.g., *Herron*, 1982; *Delmas et al.*, 1982; *Dreschhoff et al.*, 1983; *Legrand and Delmas*, 1986]. The amount of nitrate found in polar ice and snow from generally accepted background sources (biogenic decay, biomass burning, lightning, anthropogenic emissions, and stratosphere-troposphere exchange) [e.g., *Ehhalt and Drummond*, 1982; *Chameides et al.*, 1977; *Neftel et al.*, 1985; *Hameed et al.*, 1981; *Hastings et al.*, 2004] and from proposed extreme SPEs is dependent on many factors that tend to vary in an annual cycle (e.g., source strength, atmospheric processes, and transport mechanisms to the ice).

The impulsive nitrate enhancement definition used by *McCracken et al.* [2001a,b] was an extremely short duration nitrate impulse that was at least three standard deviations above and superimposed on the temporal, highly variable nitrate background. The smoothed temporal profile was subtracted from the short-duration impulse event amplitude to obtain the additional accumulated NO(y) contained in the impulsive nitrate event.

An illustrative example is the impulsive nitrate event contained in samples 1133-1136 of the GISP2-H data shown in Figure 1 that is time associated with the 23 February 1956 solar cosmic ray ground-level event [*Meyer et al.*, 1956; *Pfotzer*, 1958; *Shea and Smart*, 2012]. This is the largest relativistic solar cosmic ray event observed in the history of cosmic ray measurements with relativistic solar particles impacting the Earth's polar regions for more than 24 hours (*Švestka and Simon*, 1975). The duration of this impulsive nitrate deposition event in the ice was about 2 cm water equivalent at Summit, or 1 month.

## 3.    Data Comparisons

The data used in this analysis are from four ice cores identified by the following names: Zoe, D4, GISP2-H, and BU. The Zoe, GISP2-H and BU cores were obtained at Summit (72.6°N, 38.3°W, 3216 m elevation, 24.6 cm average annual water equivalent accumulation [*Meese et al.,* 1994; *Albert and Schultz,* 2002]), and the D4 core was drilled 237 km away in central Greenland (71.4°N, 44.0°W, 2713 m elevation, 41.4 cm average annual water equivalent accumulation [*Banta and McConnell*, 2007]). E. W. Wolff and J. R. McConnell (personal communication, 2012) have generously provided their Zoe and D4 ice core data. GISP2-H [*Dreschhoff and Zeller*, 1994; 1998] and BU [*Kepko et al.*, 2009] data were provided by G. A. M. Dreschhoff (personal communication, 2000) and L. Kepko (personal communication, 2009) respectively. Figure 2 presents the 14 year time series of nitrate concentrations from 1937 to 1951 for all four cores. The nitrate profiles look very similar when plotted in terms of water equivalent depth and are not shown here. The Zoe and D4 curves obtained from continuous flow analysis (CFA) of multiple species actually show coarser effective resolution compared to GISP2-H and BU, despite the finer resolutions listed in Table 1 of *Wolff et al.* [2012].

Discretely sampled and co-located nitrate and conductivity values were identified each 1.5 cm of firn or ice along the entire GISP2-H core yielding approximately 20 discrete samples per year (~1.25 cm water equivalent per sample). For the BU core, *Kepko et al.* [2009] developed a simplified CFA procedure that concentrated on obtaining only co-located nitrate and conductivity measurements through a single flow line, resulting in an approximately 0.063 cm water equivalent sampling interval, thus reducing dispersion effects and achieving a sampling frequency in excess of 284 (cycles) year$^{-1}$ (per year) for 1937 to 1951. To ensure the accuracy and repeatability of the data acquired, Kepko et al. also divided the initial core into four parallel sections and analyzed each section independently before averaging. This resulted in an ultrafine-resolution nitrate and conductivity data set that could be dated using the 1947 Hekla eruption volcanic tie point, annual layer counting, and the consensus Summit depth/time profile [*Meese et al.*, 1994].

In contrast, the multi-species CFA conducted on the Zoe and D4 cores used sampling intervals ranging from 0.4 to 0.9 cm water equivalent. While this CFA design enabled many geochemical species in the ice cores to be characterized in parallel, this approach, coupled with dispersion effects, inherently sacrificed effective depth resolution [*Breton et al.*, 2012] and reduced the time resolution of the Zoe and D4 data compared to those of GISP2-H and BU as shown in Figure 2.

### 3.1 Power spectral analysis

Nearly any mathematical function can be represented as the sum of a series of sinusoids; examining the nature of these has become a standard method in many fields, in this case time and depth series. For an introduction, see *Muller and MacDonald* [2002] and *Melott and Bambach* [2011]. The power spectrum is a way of showing the strength of fluctuations at various frequencies. In order to represent a sudden change over time T, frequencies up to $1/(2T)$, the Nyquist frequency, must be included. When considering a depth interval D, an analogous Nyquist frequency $1/(2D)$ is obtained. Smoothing data, which involves averaging over neighboring points, may happen by mixing the data at different times or depths. It is equivalent to removing high frequencies from the data, which can be seen in the power spectrum. The Lomb-Scargle Method [*Scargle*, 1982] was developed to analyze irregularly spaced data directly without interpolation. It is equivalent to doing a least-squares fit to sine waves of various wavelengths [*Cornette*, 2007].

In Figure 3 we show the results of a Lomb-Scargle analysis of the four nitrate data sets in terms of time. A spectral analysis of all four data sets was also obtained in terms of water equivalent depth. The depth power spectra gave similar results and are not shown, but the resulting depth resolutions are reported here. Significance levels are assigned assuming a "red noise" AR(1) background. All the data in Figure 3 show a peak near 1, corresponding to the annual nitrate cycle in polar ice that is well preserved at Summit. The discretely sampled GISP2-H data show significant power right up to the largest sampled frequency. In contrast, the CFA-sampled Zoe, D4, and BU data show a distinct drop of significance at the higher frequencies, far below the putative Nyquist frequency. We recognize that there may be some mild nonlinearities in the correspondence between time and depth scales, but these will average out over the 14-year interval with the result that there is some mixing of power for adjacent frequencies. In any case, shorter depth intervals will correspond on average to shorter times and higher frequencies. The narrowness of the peak at 1 year[-1] sets a limit of a few percent in the variation in annual snowfall rate.

The Zoe and D4 nitrate data show that even at the 50% confidence level, there is no significant power at frequencies greater than about 3 year$^{-1}$. The depth analysis of the observed nitrate pattern in the Summit Zoe data corresponds to a depth resolution that is closer to 8 cm water equivalent rather than 1 cm water equivalent, suggesting that these are at best seasonal features (i.e., the annual nitrate cycles) that are being resolved. The actual depth resolution in the D4 ice core nitrate data from 1937 to 1951 with power exceeding the 95% confidence interval is no better than 9.4 cm water equivalent at Summit. In contrast, the GISP2-H nitrate data show significant power at the 99.9% confidence level up to about 12 year$^{-1}$ (1 month period) and have a 1.9 cm water equivalent depth resolution, and the more finely depth sampled BU core nitrate data show significant power at the 99.9% confidence level up to about 36 year$^{-1}$ (~10 days period or ~0.67 cm water equivalent depth resolution). This suggests strongly but does not prove that there is a meaningful signal in the Summit ice at about a 10-day resolution that has not been consistently compromised by post-depositional processes. This resolution limit is most probably determined by the high frequency of episodic snowfall events ~88 year$^{-1}$ (~7.3 per month or roughly one snowfall event every 4 days, corresponding to ~0.28 cm water equivalent per event [*Dibb and Fahnestock*, 2004]) in combination with winds and subsequent migration of nitrate in the firn and applies to other species as well. The question we are left with then is, "What is the effective sub-annual resolution of nitrates in ice cores at Summit; i.e., how fine a resolution in the ice cores at Summit will yield useful information from multi-parameter (as well as possibly SPE/nitrate deposition) studies?"

## 3.2 Sampling frequency and data resolution analysis

It is not sufficient to show the presence of high frequencies, as in the GISP2-H and BU core power spectra (Figure 3). Noise in the data could introduce high-frequency power. In order to demonstrate that they show fine-resolution sampling, power must be correlated in such a way as to generate high peaks on a corresponding fine depth or time scale. In Figure 4 we show the five highest nitrate peaks for the interval 1937-1951 in each of these data sets. The full width at half-maximum (FWHM) was calculated by plotting each peak out to its first local minimum on either side. Peak heights and half maximum widths were plotted and calculated relative to the average nitrate concentration of each pair of local minima as the zero point. An average FWHM (FWHM$_{av}$) was then calculated from all five peaks for each core. The FWHM$_{av}$ of each core corresponds closely to the resolution limits given by the power spectra in Figure 3, but we note that the D4 core FWHM$_{av}$ of 3.5 year$^{-1}$ (3.4 months) is slightly better than the 2.7 year$^{-1}$ (4.5 months) resolution indicated by the power spectrum analysis. We therefore have good overall agreement and an independent confirmation of the resolution in the nitrate data sets.

## 4. Discussion

Summit is an optimal location for preserving the seasonal geochemical depositional record including nitrates because the relatively high snow accumulation of 0.65 m year[-1] at the surface (0.246 m year[-1] water equivalent) [*Meese et al.*, 1994; *Albert and Schultz*, 2002] is not dominated by wind erosion drifting features or sastrugi, and post-depositional processing effects are minimal [*Hastings et al.*, 2004]. Summit is also not subject to katabatic winds that re-arrange the snow at down-slope locations. Furthermore, and most importantly, any mixing of episodic snowfall events by winds will on average have the effect of reducing nitrate spike amplitudes relative to the background.

In a major study of nine soluble ions involving multiple snow pit and daily surface samples covering three years, *Dibb et al.* [2007] concluded that at least on a seasonal basis "...post-depositional changes do not greatly impact the glaciochemical records preserved at Summit." *Dibb et al.* [2007] went on to say that nitrate loss due to volatility "...is a minor ($\leq 5\%$ up to perhaps 25%) fraction of the total that is deposited..." Their findings are consistent with a previous surface and snow pit study focused on nitrates by *Burkhart et al.* [2004], who also observed impulsive nitrate features within one year of deposition at the depth and time scales discussed here and concluded that, "Over 90% of the measured nitrate in the surface snow is preserved in the pits indicating that photochemistry and temperature-dependent uptake and release from the snow do not play as important a role with preservation at Summit…" While the *Burkhart et al.* study did not observe a clear annual cycle and a consistent preservation of impulsive nitrate events, their nitrate study was limited to one year, and their results must be interpreted accordingly. These findings do not, of course, prove that impulsive nitrate features at the fine resolution indicated in the current study will be preserved at depth. However, the presence of both highly significant nitrate power spectra and consistent behavior in FWHM nitrate spikes in the BU core down to approximately 0.67 cm water equivalent depth (10 day) intervals from our results strongly support limited interference by post-depositional processes and are a validation of this analytical approach.

### 4.1 CFA dispersion effects and resolution

Mixing/dispersion effects during CFA result in coarser resolution after sampling [*Breton et al.*, 2012]. These effects are minimized in the BU core analysis and completely absent from the discretely sampled GISP2-H data as shown in Figure 2. The effects of dispersion and resolution loss on the Zoe and D4 cores can be demonstrated by (1) applying a Gaussian filter to the fine-resolution BU data of the form $\exp(-28.3t^2)$ and (2) decimating the filtered BU data by a factor of 16 (21 days) to match the sampling rate of the Zoe data from 1937 to 1951. The coefficient corresponds to a low-pass frequency filter for $f_c = 3$ year[-1] (4 months), the point at which the contribution from adjacent samples falls off from 1 at $t = 0$ to $\exp(-\pi)$. The filtered/decimated

BU data and their power spectrum are presented in Figures 5a and 5b, respectively. Filtering the BU data to the Zoe resolution simulates how mixing in the CFA fluid stream smoothes and removes the high-frequency peaks as shown in Figure 5a. The resulting power spectrum of the filtered BU data in Figure 5b strongly resembles that of Zoe shown in Figure 3a.

### 4.2    Known extreme high-energy solar proton events in the 1940s

The first observed SPEs were the relativistic high-energy (>4 GeV) events detected by cosmic ray ionization chambers in 1942, 1946, and 1949 [*Forbush*, 1946; 1950]. Figure 6 shows all of the high-energy solar cosmic ray ground-level events observed by ionization chambers and/or muon telescopes since the beginning of the continuous cosmic ray measurement era (1936 to present). Such events are expected to ionize the atmosphere down to low altitudes [*Atri* et al., 2010; Usoskin et al., 2011]. As can be seen in Figure 6, the events in the 1940s were extremely large over a short time period compared to events during the contemporary spacecraft era (1965 to present). In an examination of the BU core in Figure 2d, the large-amplitude impulsive nitrate increases (more than 10 standard deviations above the pre-event background) in the core are time associated with the first known very high-energy solar cosmic ray events (28 February and 7 March 1942 (combined); 25 July 1946, 19 November 1949). This ultrafine-resolution CFA method resulted in a FWHM of 9 days (0.6 cm water equivalent) for the impulsive nitrate increases coincident with these high-energy relativistic solar cosmic ray events. The probability that during a 14-year period (i.e. 168 months) you would purely by chance select three events that would correspond this closely to the month of known high-energy solar cosmic ray events (February/March, 1942, July 1946, and November 1949) is very small ~$(3/168)^3$, or about 1 in 175,000. Nitrate increases are also observable in the GISP2-H core but not as distinctly as in the higher-resolution BU core. None of these events is uniquely discernible in the Zoe or D4 cores shown in Figure 2.

Figure 7 presents the BU (nitrate and conductivity) and the Zoe (multi-species) data on the same time scale. This is an interval when the annual cycles for these two cores (BU and Zoe) are in good synchronization, based on the surface and 1947 Hekla eruption tie points, the *Meese et al.* [1994] depth-age scale, and interpolation using species that display annual cycles; the data are plotted directly as acquired without any adjustments. Despite differences in resolution, note that these impulsive nitrate increases found in both the BU and GISP2-H cores do not correspond in time to indicators of biomass burning such as anomalous increases in ammonium or black carbon in the Zoe and D4 multi-species data such as shown in Figure 7.

In the BU core analysis, *Kepko et al.* [2009] also identified small nitrate increases that correspond to the time of strong polar cap blackouts recorded by vertical ionospheric soundings that were identified as probable solar proton events by *Švestka* [1966]. Solar proton events undoubtedly occurred during the solar maximum years of the 17th and 18th solar cycles;

however, the only known high-energy solar cosmic ray events are those recorded by the Forbush ionization chambers. The only short-duration impulsive nitrate events that are time associated to high-energy solar cosmic ray events are the increases in 1942, 1946, and 1949. This is why we have concentrated on these well-documented relativistic solar cosmic ray events to show that the lack of temporal resolution in the D4 and Zoe cores do not permit the identification of these short-lived impulsive nitrate events.

## 5.    Conclusions

*Wolff et al.* [2012] did not find a nitrate increase in the ice cores they examined corresponding to the 1859 Carrington event. Subsequently, their analysis of this single, incompletely characterized event has been used to discredit impulsive nitrate events as possibly indicative of large fluence SPEs [e.g., *Schrijver et al.*, 2012; *Cliver et al.*, 2014]. After examining the data provided by Wolff et al. that overlap the GISP2-H and BU data for the well-documented, large, high-energy relativistic solar cosmic ray events in the 1940s, we conclude that their data resolution was inadequate to differentiate any of the impulsive nitrate events in the test period of 1937-1951. The limitations on significant data resolution mean that the Zoe and D4 data sets cannot confidently detect or represent changes at frequencies greater than ~3 year$^{-1}$, or on time scales of less than approximately 3-5 months. The 36 year$^{-1}$ (~10 day) average resolution that is obtainable, as indicated by the BU core, corresponds to a depth resolution < 1 cm water equivalent at Summit. These findings also imply that ice core analysis of nitrate at a resolution such as Zoe and D4 probably could not discern the Carrington event as well.

We note that the GISP2-H core power spectrum up to its sampling interval frequency appears to be similar to the BU core power spectrum as shown in Figures 3c and 3d. The fact that the BU data resolution makes it possible to discern impulsive nitrate events means that there is valuable information down to time scales of order 10 days in Summit ice core data that is currently not being exploited.

These results have relevance and impact beyond the issue of SPE resolvability in the polar ice core nitrate record. This study shows that the ice sheet at Summit contains sub-seasonal information in at least some species at the approximate time scale of snowfall events that can be extracted relatively easily for comparison with current and historical climatological records and applied to studying interactions between different species. This will allow glaciologists and atmospheric scientists to better understand sources in the atmosphere and the transport mechanisms of impurities to the ice sheet, as well as post-depositional processes.

We suggest additional fine-resolution, multiple species analysis that include nitrate and potentially interfering and "fixing" impurities (i.e., ammonium, black carbon, vanillic acid, dust, and sea salt) with replicate sampling at times of other potentially large-amplitude solar events in the last two millennia (e.g., [14]C and [10]Be events) such as events documented circa 774-775 A.D. and 993-994 A.D. [*Miyake et al.*, 2013; *Melott and Thomas*, 2012; *Usoskin et al.*, 2013]. A similar recommendation was made by *Wolff et al.* [2008]. We stress that astrophysical ionization events are not expected to produce any of the chemical species attributable to volcanoes or wildfires, other than nitrate. This is important, as there are indications of much more energetic events on stars similar to the Sun [*Nogami et al.*, 2014] which are potentially very dangerous if they occur on the Sun as well. The medieval events were very high in fluence or had extreme spectral hardness since [14]C is dependent on high-energy proton fluence [*Thomas et al.*, 2013], and ice core nitrate studies may be important in our understanding of these events by effectively giving information on the total energy deposition in the atmosphere, and thus a crude approximation to the spectrum and total energy deposition. Soft solar events with enough fluence to account for the observed [14]C in 774-775 A.D. and 993-994 A.D. can severely deplete the ozone layer and are potentially catastrophic [*Thomas et al.*, 2013].

**Acknowledgements.** We appreciate the detailed Zoe and D4 data from E. W. Wolff and J. R. McConnell. These data are available at [http://www.ncdc.noaa.gov/paleo/icecore/Greenland/Greenland.html](http://www.ncdc.noaa.gov/paleo/icecore/Greenland/Greenland.html) . We also acknowledge G.A.M. Dreschhoff and L. Kepko for providing the detailed GISP2-H and BU core data respectively. The original GISP2-H data are available at [http://nsidc.org/data/arcss026.html](http://nsidc.org/data/arcss026.html) and in graphic form by sample number in *Dreschhoff and Zeller* [1994]. The BU data are available from L. Kepko. Data for Figure 6 are from *Shea et al.* [1979], *Smart and Shea* [1991] *Poirier and D'Andrea* [2002], and *D'Andrea and Poirier* [2005]. This research is completely unfunded. We are grateful for comments from T. P. Armstrong and K. G. McCracken and from the referees.

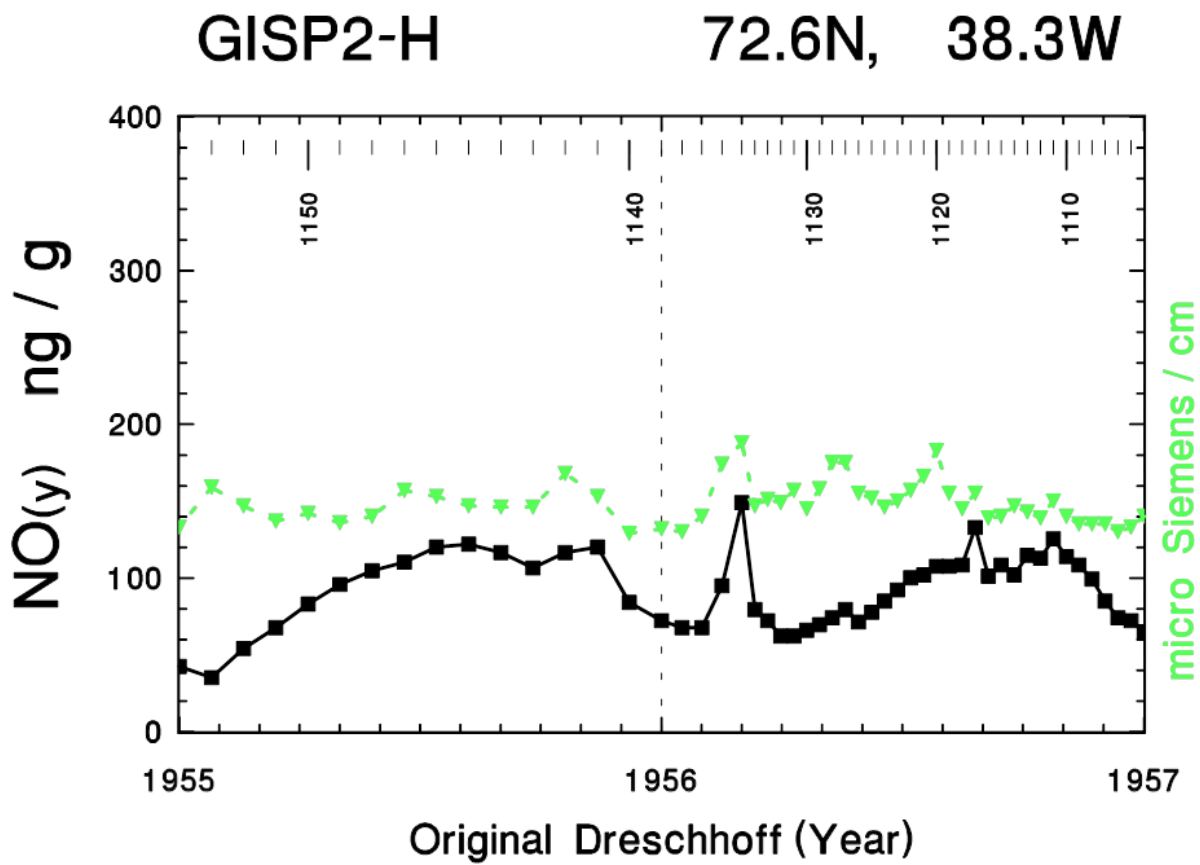

**Figure 1.** Illustration of the nitrate (squares) and conductivity (inverted triangles) data from the GISP2-H ice core from 1955 to 1957. Notice the annual NO(y) cycles (summer high - winter low). The superimposed impulsive NO(y) event identified by *McCracken et al.* [2001a] is in samples 1133-1136 (a 34-day interval) as indicated by the sample numbers at the top of this figure. Each sample represents 1.5 cm of deposited snow or consolidated firn. The small impulsive event in August 1956 appears likely to be correlated with the small ground-level (neutron monitor observed) SPE on 31 August 1956 [*McCracken*, 1959].

.

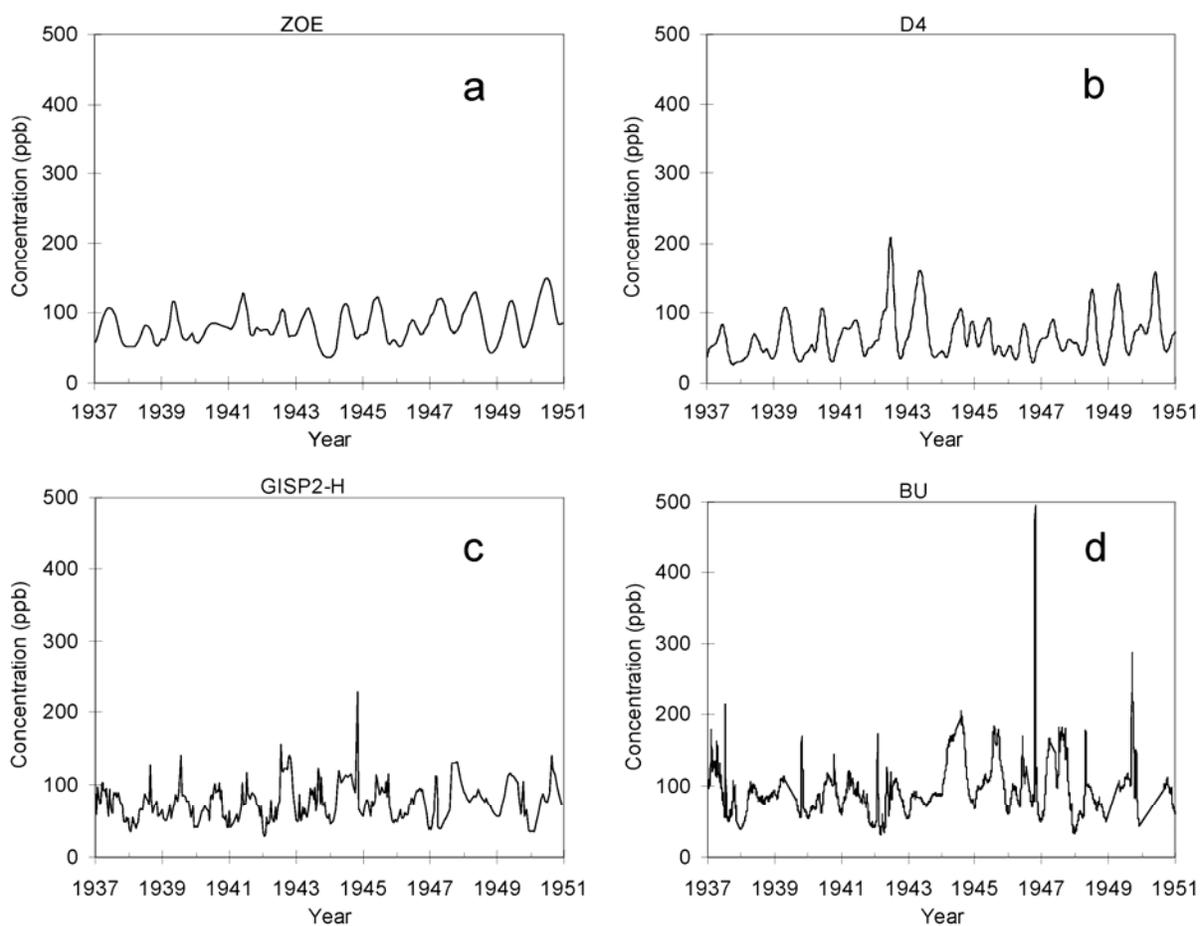

**Figure 2.** The time series of nitrate concentration fluctuations for (a) Zoe, (b) D4, (c) GISP2-H, and (d) BU ice cores from 1937 to 1951. Note the progressive improvement in resolution, as evidenced by more and shorter-term nitrate events, when passing through these series from Figures 2a-2d.

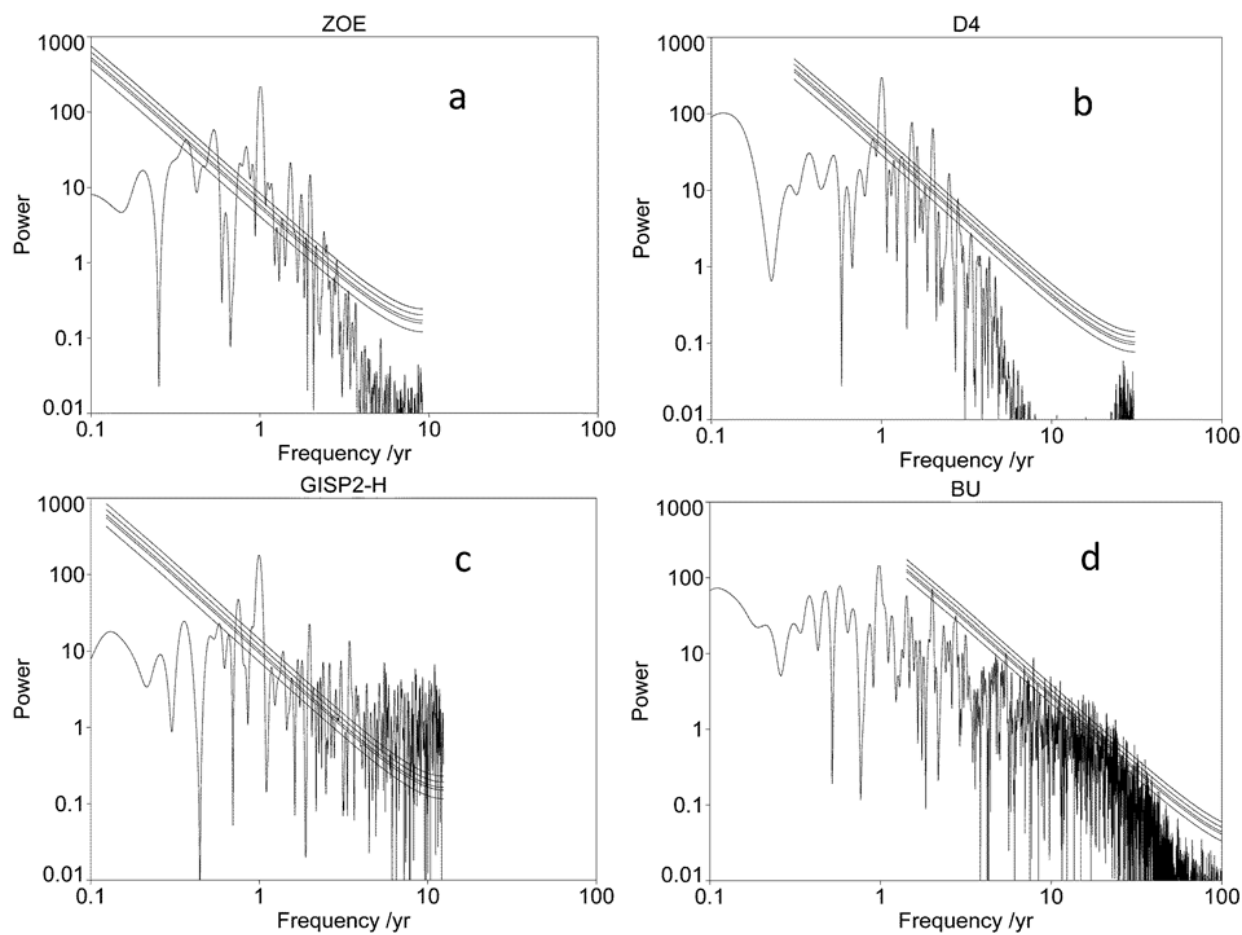

**Figure 3.** The power spectra of nitrate concentration fluctuations for (a) Zoe, (b) D4, (c) GISP2-H, and (d) BU data sets. All are shown up to the maximum sampled (Nyquist) frequency (in cycles per year) except BU, which is about 140 year$^{-1}$, where one year represents, on average, about 24.6 cm of water equivalent deposition at Summit. The significance levels (the five parallel lines) are computed by the software using an AR(1) background as the probability of finding the highest peak in the curve above them, for significance levels from bottom to top of 50%, 10%, 5%, 1%, and 0.1% (respectively). Note that Zoe and D4 show drop-offs in power that are no longer statistically significant at even the 50% confidence level at about 3 year$^{-1}$. This indicates that the Zoe and D4 nitrate data cannot resolve significant information on time scales less than about 4-5 months (~8 cm water equivalent at Summit). Significance levels stop in BU at 1.4 year$^{-1}$ due to a two order-of-magnitude limit in the Autosignal version 1.7 software and are affected by the artificially high frequencies present. If these frequencies are truncated, the significance lines move downward, so the plotted levels represent the most conservative choice.

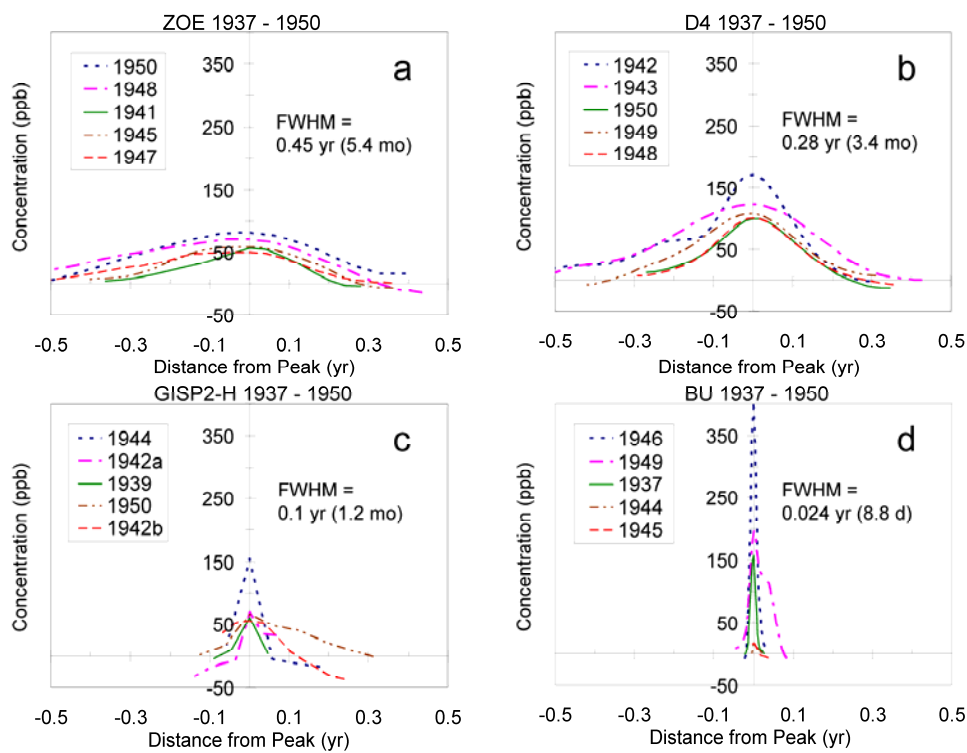

**Figure 4.** Five highest ice core nitrate concentration peaks and FWHM$_{av}$ covering 1937-1951 for (a) Zoe 5.4 months (11 cm water equivalent), (b) D4 3.4 months (12 cm water equivalent at the D4 site, 6.9 cm water equivalent at Summit), (c) GISP2-H 1.2 months (2.5 cm water equivalent), and (d) BU 8.8 days (0.6 cm water equivalent). This method suggests that the resolution limits for the D4 and Zoe nitrate data are 3-5 months, in agreement with the spectral analysis. See text for details.

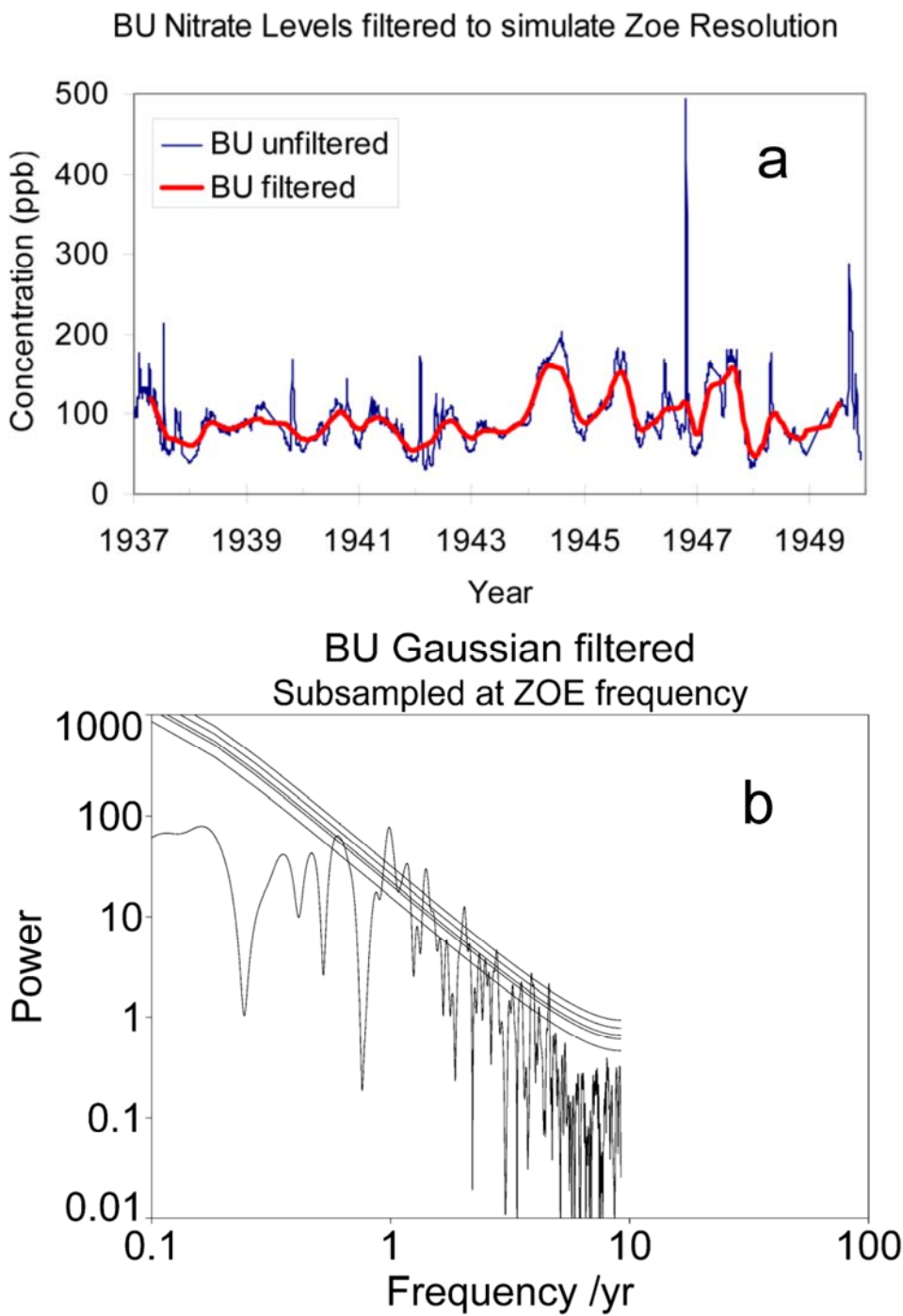

**Figure 5.** (a) The BU data are shown in original form and also Gaussian-smoothed and decimated consistent with the power spectrum of Zoe as shown in Figure 3 (see text). (b) The power spectrum of the modified data is confirmed to resemble that of Zoe in Figure 3a. Impulsive events resolved in the unfiltered BU data in Figure 5a are not resolved when the data emulate Zoe.

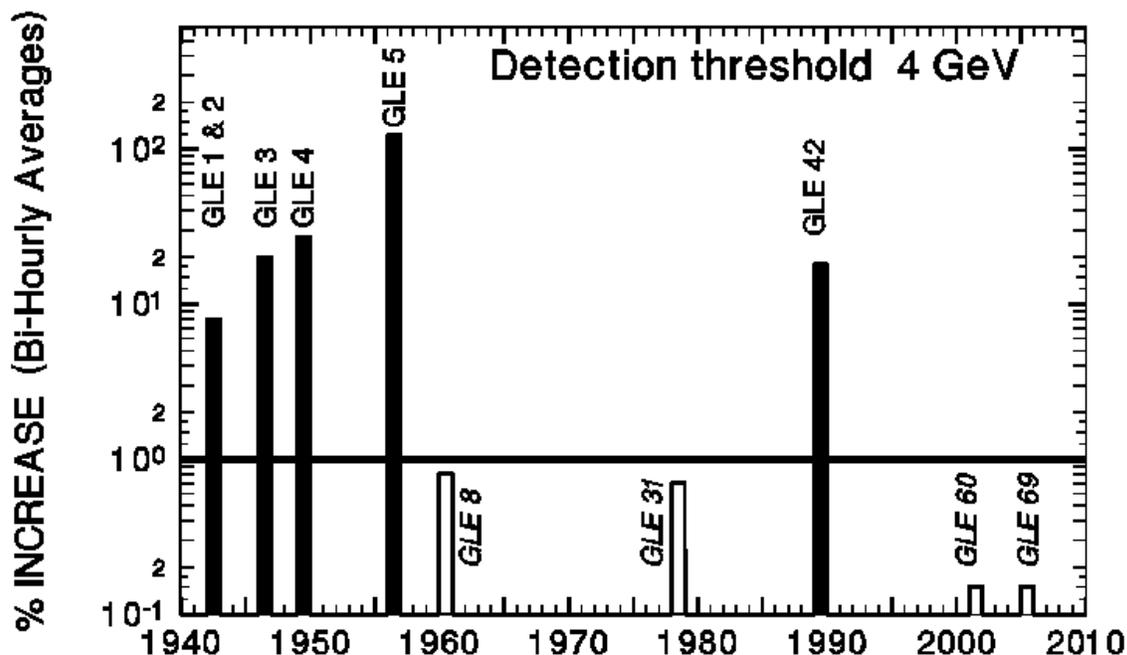

**Figure 6.** The high-energy solar cosmic ray events recorded by muon detectors from 1940 to 2010. The magnitude of the increases after 1950 have been normalized to bi-hourly data equivalent in order to be consistent with the data reported from the ionization chambers in the 1940s. The ultra-sensitive muon detectors constructed after 2000 are capable of resolving very small events. The cosmic ray community identifies ground-level events (GLEs) by a sequence number from the first recorded GLE. The dates for the GLEs shown in this figure are the following: GLE 1 (28 February 1942), GLE 2 (7 March 1942), GLE 3 (25 July 1946), GLE 4 (19 November 1949), GLE 5 (23 February 1956), GLE 8 (4 May 1960), GLE 31 (7 May 1978), GLE 42 (29 September 1989), GLE 60 (15 April 2001), and GLE 69 (20 January 2005). The small GLE muon events (<1% increase) probably do not have enough flux to generate an impulsive NO(y) event.

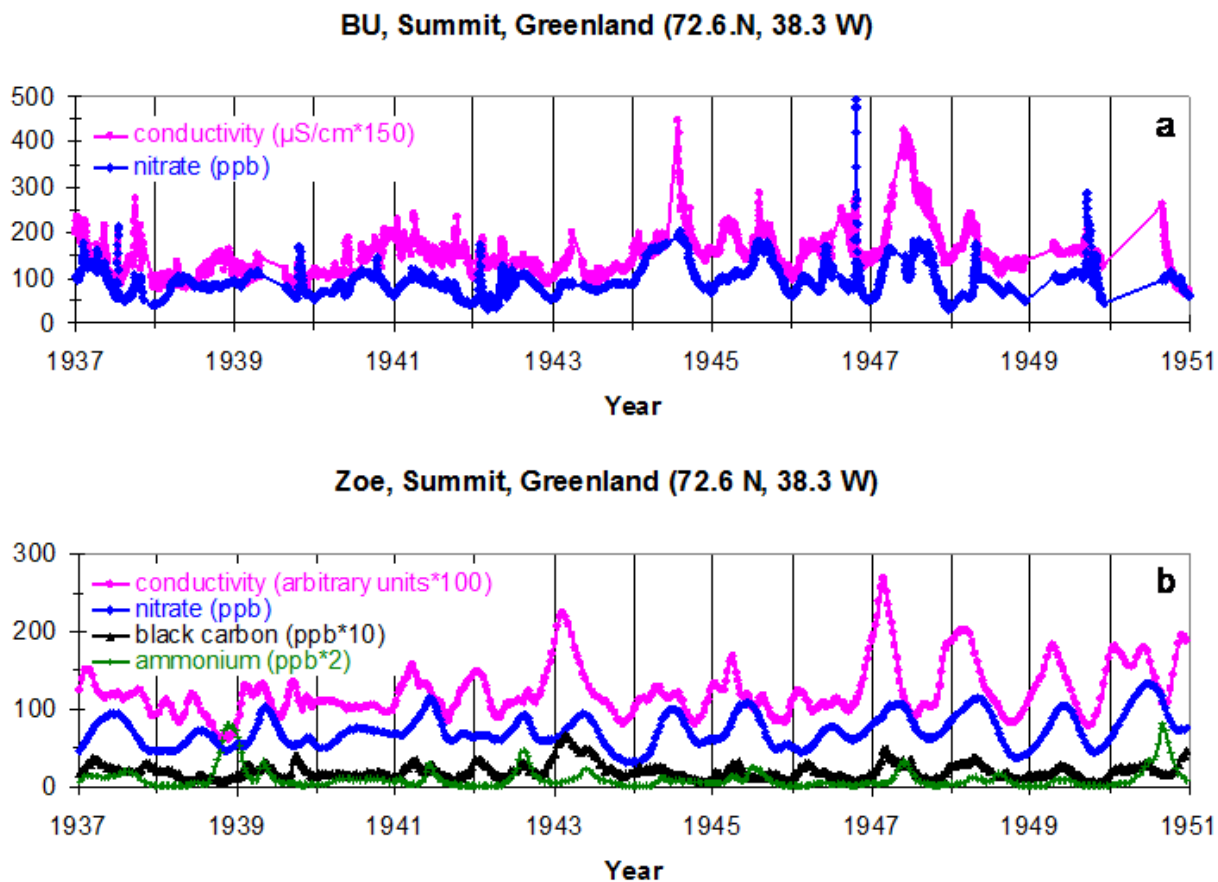

**Figure 7.** (a) BU nitrate and conductivity data, and (b) Zoe multi-species data. The timing in the BU data is synchronized to the large conductivity enhancement (also observed in the Zoe core) that persists for several months corresponding to the eruption of the Hekla, Iceland, volcano on 29 March 1947. The ammonium data have been multiplied by 2 and the black carbon data multiplied by 10 to illustrate the time periods when biomass burning could make a significant contribution to the nitrate signal. Note that the impulsive nitrate events found in the BU core in late 1946 and late 1949 do not correspond in time to the seasonally resolved (mostly summer) indicators of biomass burning such as ammonium or black carbon found in the Zoe multi-species data, as shown here in the winters of 1938-1939 and 1942-1943 and the summers of 1947 and 1950.